\documentclass[twocolumn,preprintnumbers]{revtex4}
\usepackage{graphicx}
\usepackage{dcolumn}
\usepackage{bm}
\begin{document}

\preprint{}

\title{Self-organized quantum transitions in a spin-electron coupled system}

\author{W. Koshibae$^{1}$}
\author{N. Furukawa$^{2,3}$}
\author{N. Nagaosa$^{1,4}$}
\affiliation{%
$^1$Cross-Correlated Materials Research Group (CMRG), RIKEN, Saitama 351-0198, Japan
\\
$^2$Aoyama-Gakuin University, 5-10-1, Fuchinobe, Sagamihara, Kanagawa 229-8558, Japan
\\
$^3$ERATO-Multiferroics, JST, c/o Department of Applied Physics, The University of Tokyo, Tokyo 113-8656, Japan
\\
$^4$Department of Applied Physics, The University of Tokyo, Tokyo 113-8656, Japan
}%
\begin{abstract}
We investigate quantum dynamics of the excited electronic states 
in the double-exchange model at half-filling by solving coupled 
equations for the quantum evolution of electrons and 
Landau-Lifshits-Gilbert equation for classical 
spins. The non-adiabatic quantum transitions driving the 
relaxation are coordinated through the self-organized space-time 
structure of the electron/spin dynamics leading to 
a resonant precession analogous to the ESR process.  
\end{abstract}

\pacs{PACS numbers: }
\maketitle
In the strongly correlated electron systems such as the transition metal oxides,  
the rich phases including the spin/charge/orbital ordered states 
and superconducting state are realized by the electron-electron 
and electron-lattice interactions~\cite{Imada}. 
An essential feature of the strongly correlated electron systems is the 
appearance of the internal degrees of freedom such as spins.
The dynamics of the electrons with the strong interaction   
can be translated to the 
motion of the electrons in the background of the 
fluctuating {\it spin fields} via the Stratonivich-Hubbard 
transformation \cite{Imada}.
Therefore, the quantum dynamics of the correlated electrons can be 
formulated as that of the coupled dynamics of the spins and electrons.

In particular, the time-dependent 
quantum dynamics in nonequilibrium states are now 
an issue of intensive interests.
An example is the photo-induced quantum dynamics of the spin-electron coupled 
systems \cite{spincross,munekata,koshihara,miyano,fiebig}.   
In the systems, the dynamics of the spins plays a crucial 
role in the relaxation of the electrons after the photo-excitations
leading to distinct features compared with the usual semiconductors
or band insulators, where the electron-phonon interaction
plays the major role in the relaxation. 
For example, the photo-induced ferromagnetism 
can leads to the metallic state in manganites due to 
enhanced kinetic energy in the double exchange 
interaction \cite{miyano,fiebig} in sharp contrast to 
the self-trapping of the small polaron in the band insulators. 
This means that the coupled dynamics of the spins and electrons 
must be treated as the many-body collective phenomena in the 
correlated electron systems.
The nonequilibrium quantum phenomena still remain challenges 
for theoretical study~\cite{matsueda1,matsueda2,yonemitsu}.
In general, it is not easy to capture the nature of 
the many-body electronic state, 
and our study below is complementary to the previous works 
giving a semiclassical picture for quantum dynamics
for larger size/higher dimensional systems.

In this paper, we study the relaxation dynamics of the excited states
in the double-exchange model~\cite{deGennes}, 
where the classical spins are coupled to 
the conduction electrons. This model has the advantage that the 
fully quantum dynamics of the electrons combined with the classical
motion of spins can be simulated, providing a clear physical
picture. This way of describing the correlation effect is essentially 
justified for the magnetically ordered state where the (staggered)
magnetization behaves as semi-classical object.
We found that the self-organized
space-time structure of the spins and electrons is formed through the
{\it quantum} transition as shown below.
 
We start with the Hamiltonian on the square lattice, 
\begin{eqnarray} 
\label{Hamiltonian}
\hat H=-t\sum_{<ij>,s}c^\dagger_{is}c_{js}+h.c.
-J_H\sum_{iss'}
c^\dagger_{is}c_{is'}\vec{\sigma}_{ss'}
\cdot \vec{S}_i,
\end{eqnarray} 
where $<$$ij$$>$ denotes a nearest-neighbor pair, $s$ and $s'$ 
are indices for electron spin, respectively, 
and $\vec{\sigma}_{ss'}$ is given by Pauli matrices. 
The local spins, $\vec{S}_i$'s, are taken to be classical vectors with 
magnitude $S$. Other notations are standard.  
We consider the half-filled case, i.e., one electron per site.  

Using finite size systems, we numerically investigate 
the time evolution of the  electronic states and local spins. 
The equation of motion for the local spins is expressed by the 
Landau-Lifschitz-Gilbert (LLG) equation, 
$
\dot{\vec{S}}_i=
-J_H\langle\vec{\sigma}_i\rangle\times\vec{S}_i
+\alpha\vec{S}_i\times\dot{\vec{S}}_i,
$
where $\langle\vec{\sigma}_i\rangle$ is the expectation value of electron spin 
at site $i$, and the Gilbert damping constant $S\alpha$ 
describes all the other relaxation processes 
of the spins than the coupling to the conduction electrons.  
When $\langle\vec{\sigma}_i\rangle$ is fixed, the solution of the LLG equation is given by
$
\phi_i(T)=-[(SJ_H/t)|\langle\vec{\sigma}_i\rangle|/(1+(S\alpha)^2)](t/S)T,
$
and 
$
\theta_i(T)=2\tan^{-1}[\tan(\theta_{i0}/ 2)\exp
(S\alpha\phi_i)],  
$
 where $(\theta_i,\phi_i)$ is the polar and azimuthal angles of 
$\vec{S}_i$ in the local spin coordinate
where $z$ axis is parallel to  
$\langle\vec{\sigma}_i\rangle$ and initial value of $\phi_i$ is zero, 
 and $T$ is time.  
The initial value of $\theta_{i}$ is written by $\theta_{i0}$.  
Evolution of the electronic state $|\Phi(T)\rangle$ is given by
$ 
|\Phi(T)\rangle=\hat U(T) |\Phi(0)\rangle
$, 
where $\hat U(T)$ is a unitary operator for the time evolution.
If  $\{\vec S_i\}$ are fixed, we have
$ \hat U(T)=\exp( - i \hat H T)$.
We successively calculate the evolution of the electronic state and the LLG equation  
for a small time increment $\Delta T$ 
to investigate evolutions
of the system in total.
The changes in  $\{\vec{S}_i\}$ and
 $|\Phi( T)\rangle$  
 are reflected
to the calculation 
through  $\hat H$ and $\{\langle\vec{\sigma}_i\rangle\}$.

The Hamiltonian (\ref{Hamiltonian}) 
is expressed by a bilinear form of 
fermion operators.  
In such a case, it is known that
the electronic state $|\Phi(T)\rangle$ remains to be a single Slater
determinant state if the initial state is so~\cite{IH}.
This allows us to increase the system size,
which is advantageous for investigation of low-frequency dynamics with
accuracies.

\begin{figure}[t]
\includegraphics[width=80mm,clip]{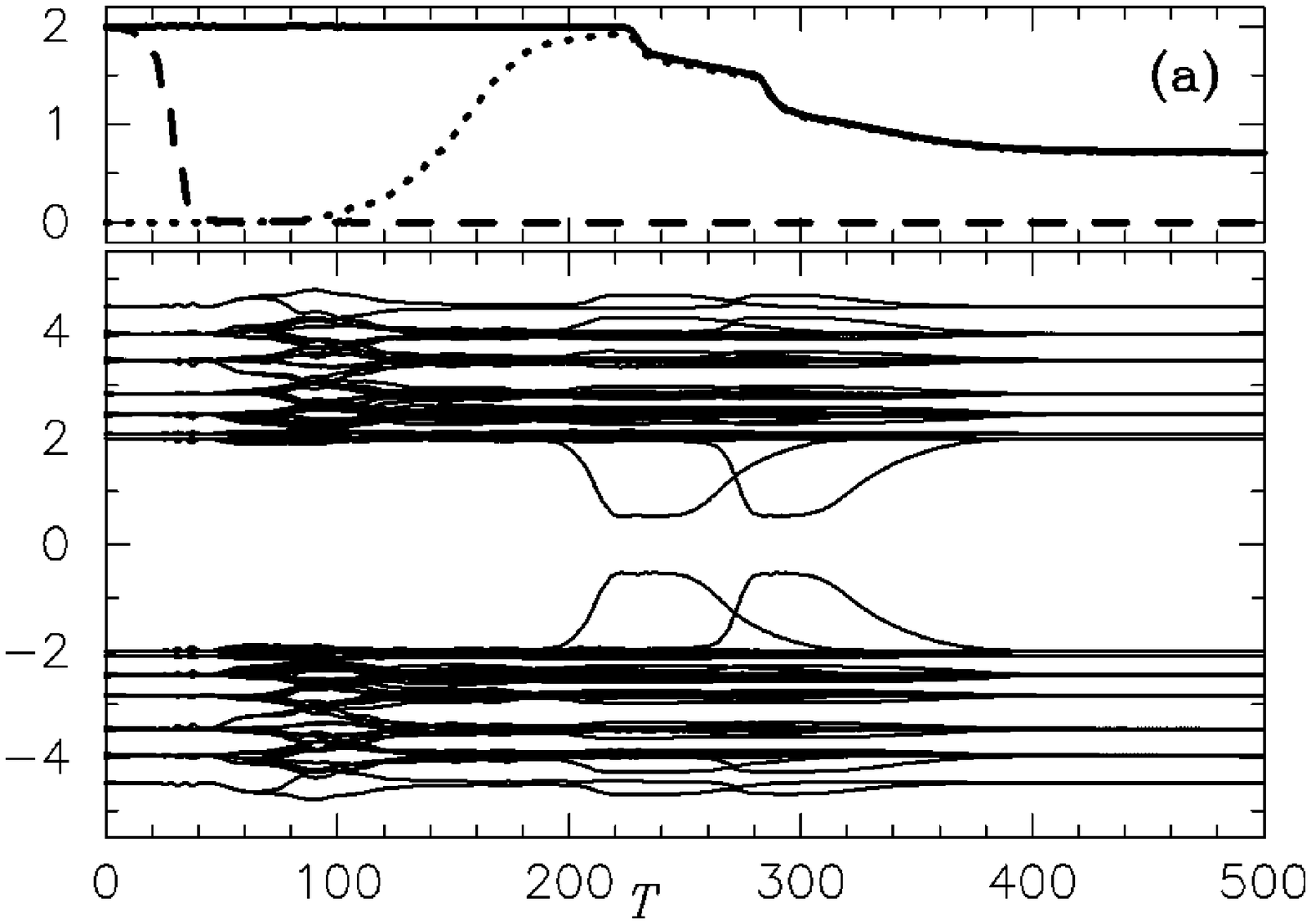}
\begin{tabular}{cc}
\begin{minipage}{0.65\hsize}
\includegraphics[width=55mm,clip]{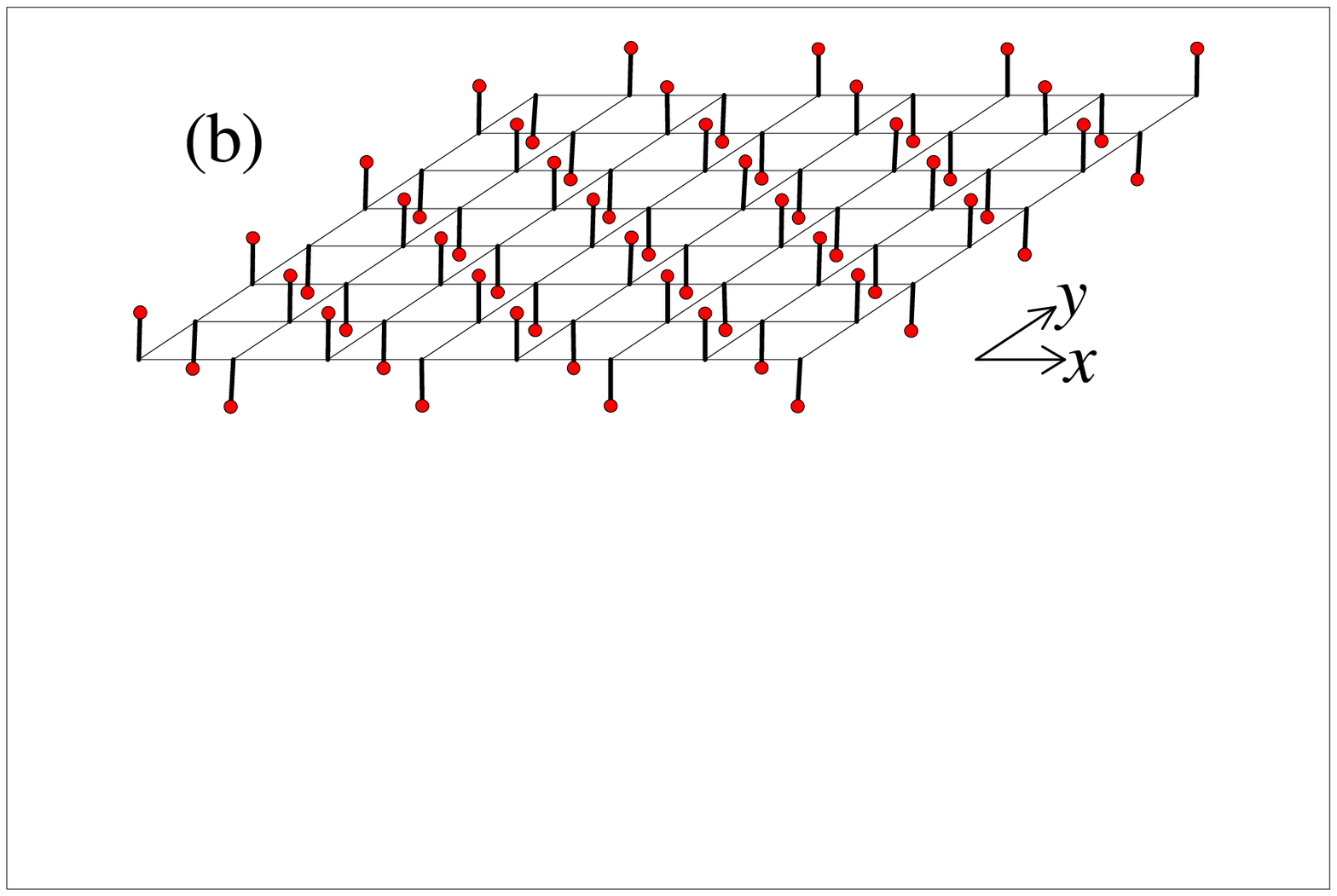}
\includegraphics[width=55mm,clip]{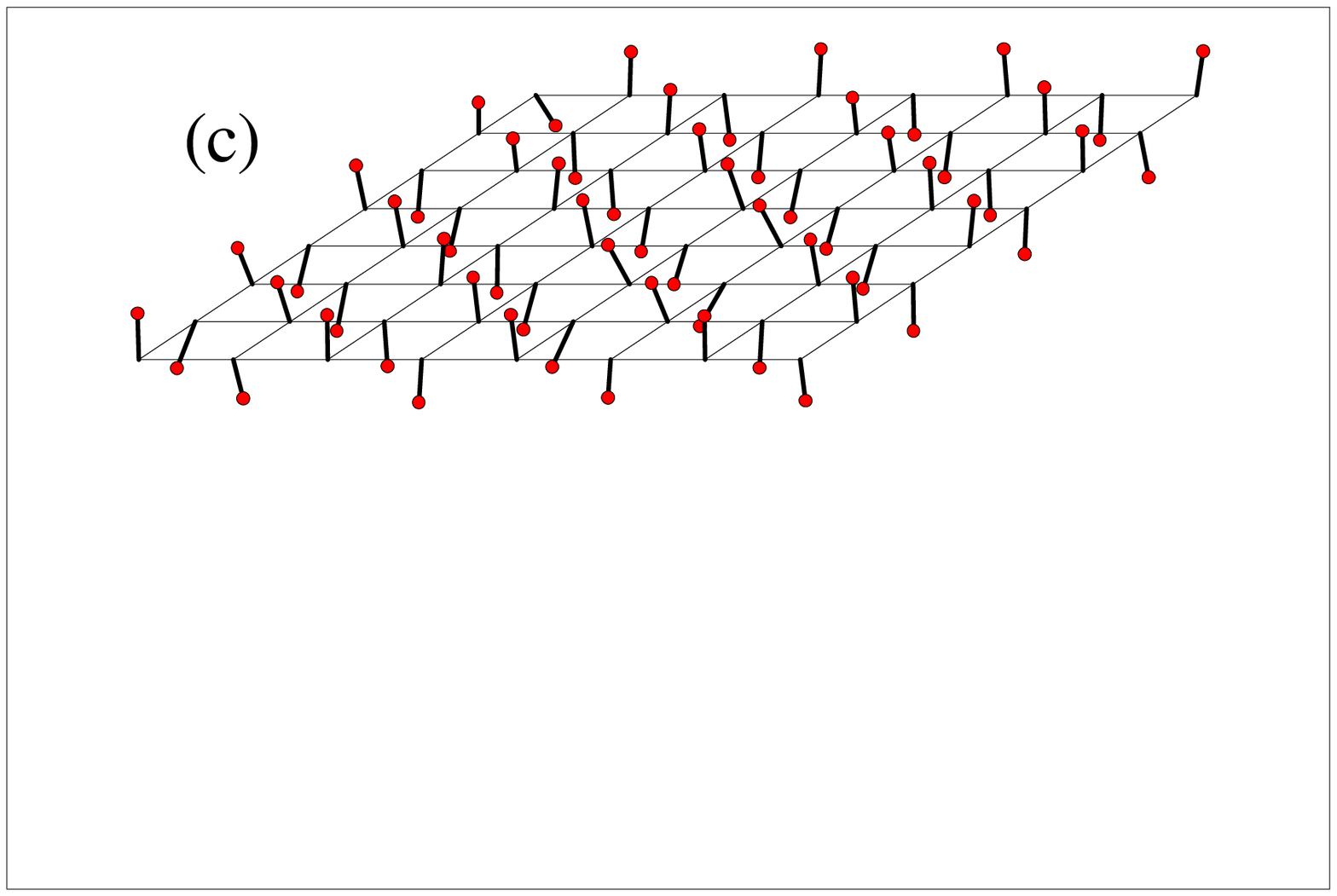}
\includegraphics[width=55mm,clip]{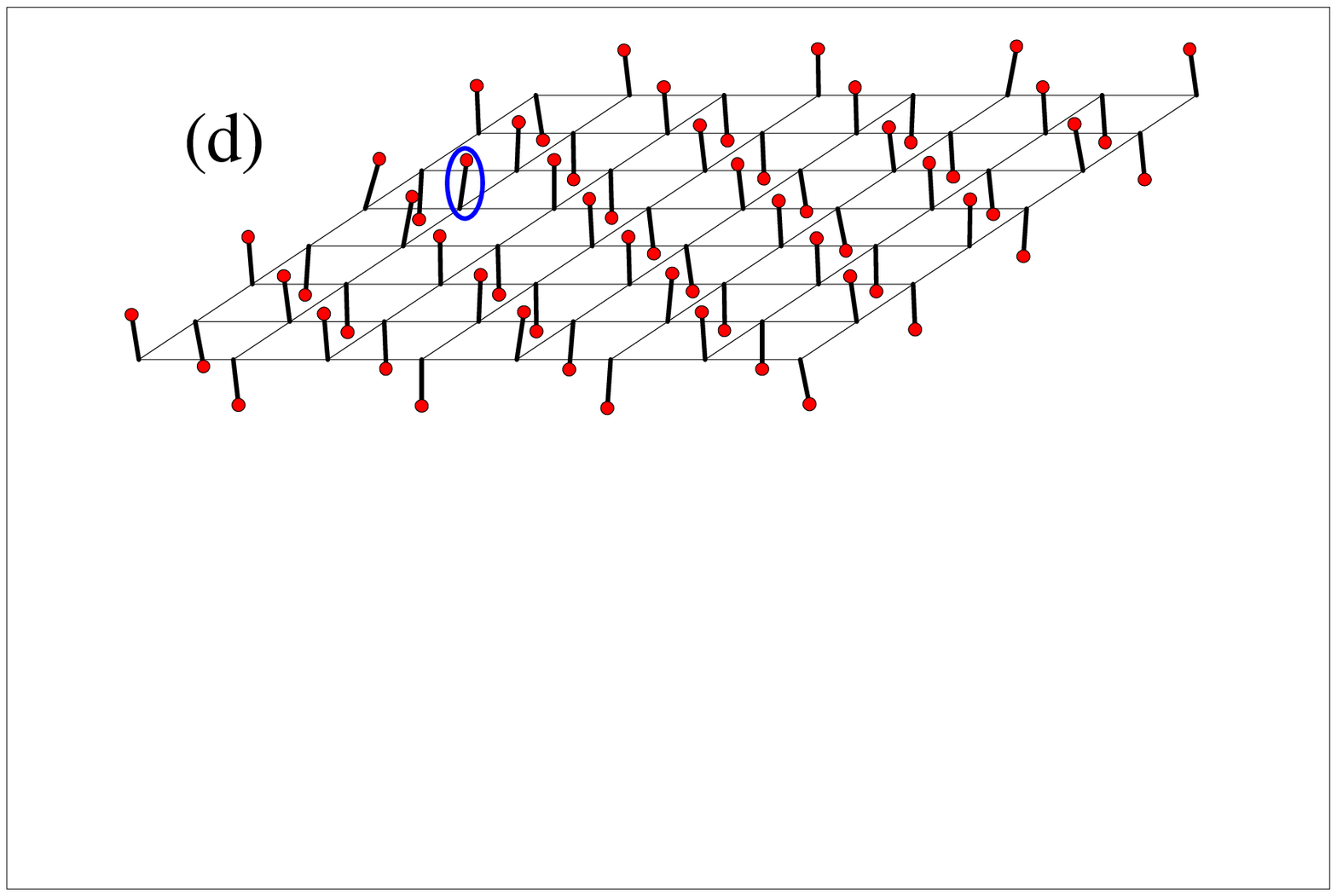}
\includegraphics[width=55mm,clip]{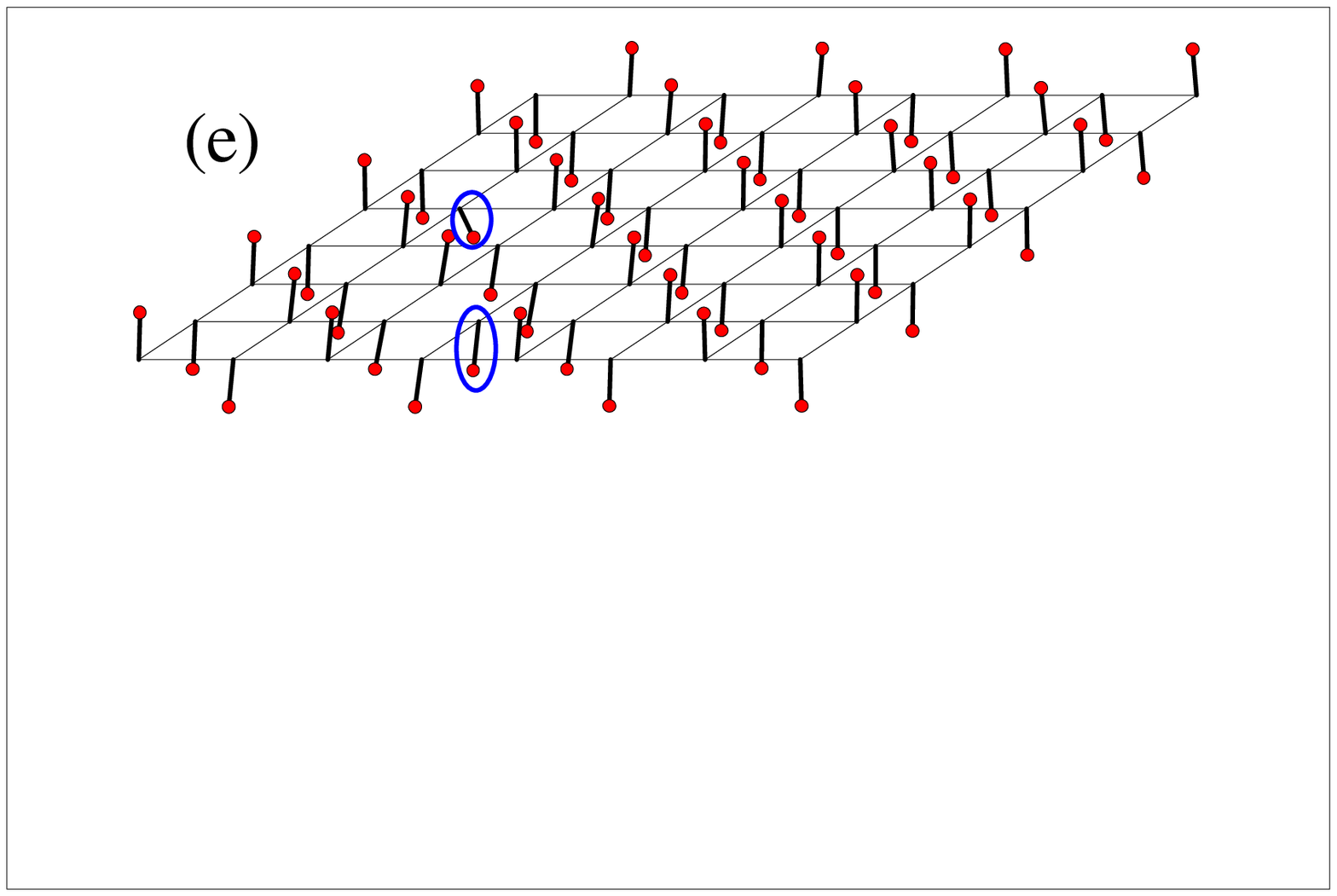}
\end{minipage}
\begin{minipage}{0.3\hsize}
\includegraphics[width=30mm,clip]{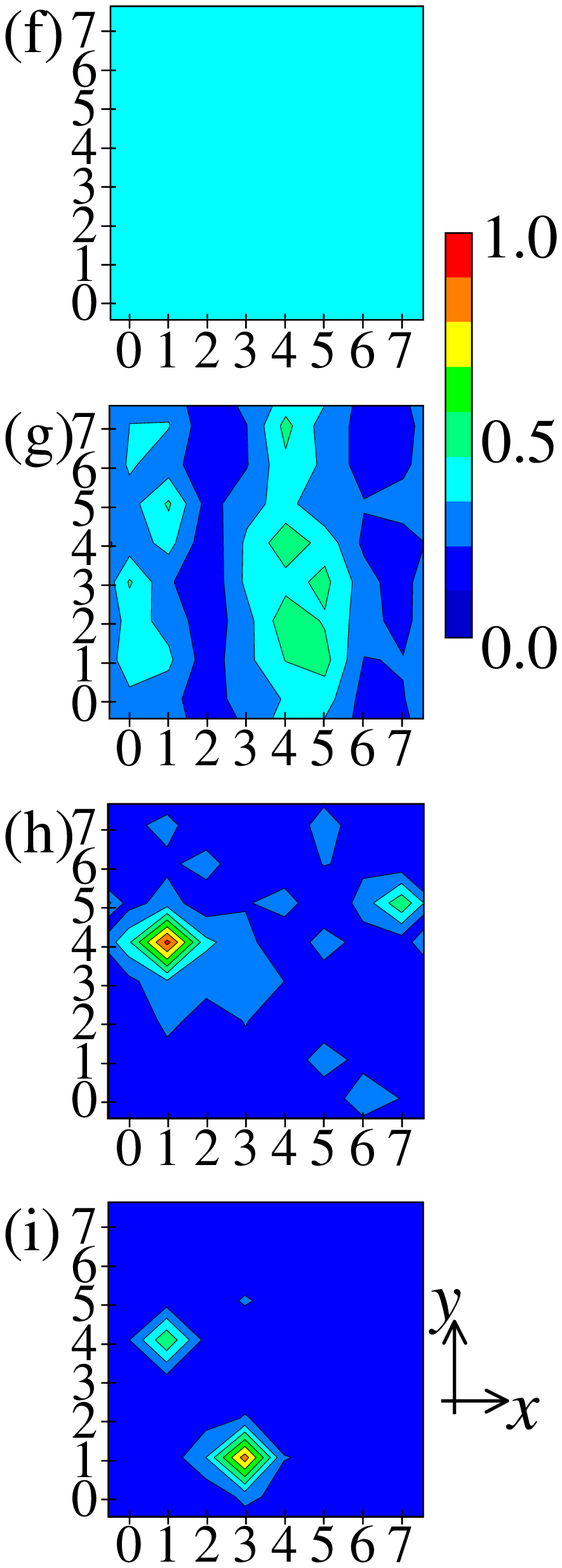}
\end{minipage}
\end{tabular}
\vspace{-10pt}
\caption{\label{jh2b}
(Color online) Time evolution of the double exchange model (see text).   
(a) The lower panel shows the time ($T$) dependence of the energy level structure.     
In the upper panel, 
the number of electrons in the upper energy band (solid line), 
in the $highest$ two energy states (broken line), 
and in the $lowest$ energy states of the upper energy band (dotted line) 
are shown.   
Configuration of local spins at $T$=8, 80, 224, and 288 are presented 
in (b), (c), (d) and (e), respectively. 
Dot indicates the head of local spin.   
The local spins marked by ellipses in (d) and (e) are strongly deviated from the 
ground state.   
Figures (f), (g), (h) and (i) show the distribution of excitation energy density 
measured from the ground state at $T$=8, 80, 224, and 288, respectively. 
}
\end{figure}

Figure \ref{jh2b} shows the calculated result 
on the system of size $8 \times 8$ with the periodic boundary
condition.
As a typical example, a parameter set, 
$t$=1, $SJ_H$=2, $S\alpha$=1, $S$=1, $\Delta T$=0.008, is used.  
The lower panel of Fig.~\ref{jh2b}(a) is the time ($T$) dependence of 
the energy level structure.   
The Fermi level is taken to be zero.  
At around $T$$\sim$0, we clearly see the energy gap $2SJ_H$ between upper and lower 
energy bands in the lower panel of Fig.~\ref{jh2b}(a).  
We prepare the initial state in the following way:  
The ground state of the double-exchange model (\ref{Hamiltonian}) at half filling 
is the antiferromagnetic (AF) insulating state because of 
the perfect nesting-condition in this system.  
In order to mimic the thermal fluctuation, 
we introduce a random tilting of each spin from the AF configuration up to 0.1 radian 
which corresponds to the state with an excitation energy of $\sim$0.001$t$ 
from the ground state.   
The energy band is divided into upper and lower ones.    
For the initial state, we fill up the lower energy band, and then 
transfer two electrons from the lowest eigenstates 
in the lower energy band to the two highest levels in the upper energy band.  
In the upper panel of Fig.~\ref{jh2b}(a), 
the number of electron of the $highest$ ($lowest$) energy states in the upper energy band 
is shown by 
the broken (dotted) line, 
and the solid line is the number of electron in the upper energy band.  
Figures \ref{jh2b}(b)-(e) show the time dependence of 
the configuration of the local spins, 
while Figs.~\ref{jh2b}(f)-(i) are the corresponding local energy density 
defined by the expectation value of  
$ -\left(t/2\right)\sum_{\rho,s}(c^\dagger_{is}c_{\rho s}+h.c.)
-J_H\sum_{ss'}c^\dagger_{is}c_{is'}\vec{\sigma}_{ss'}
\cdot \vec{S}_i,  
$  
where $\rho$ runs over the nearest-neighbor sites of $i$.  
In the Figs.~\ref{jh2b}(f)-(i), 
the energy of the ground state is taken to be zero.  

In the numerical simulation, we find several distinct time regions, 
Stage (I)-(IV):

\noindent
{\bf Stage (I)}: Precursor process.  Up to $T$$\sim$20, 
there occurs almost no change and the spins are
almost frozen at the initial configuration. 
Figure \ref{jh2b}(f) shows the energy density  
at $T$=8, showing that the excitation energy 
is almost uniformly distributed 
due to the extended nature of the excited states.  
However, small deviation from the AF configuration grows 
eventually leading to the next stage.  

\noindent
{\bf Stage (II)}: Self-organization process. At this stage, the system shows spatial  
inhomogeneity with energy dissipation.  
At around $T$=20$\sim$40, the spins start to move, and the electrons at the 
highest 2 energy states start to spread into lower energy states within the 
upper energy band (see the broken line in the upper panel of Fig.~\ref{jh2b}(a)).
Up to $T$$\sim$170, the deviation of the spins are not so large moving around the
original direction as shown by the snapshot in 
Fig.~\ref{jh2b}(c) at $T$=80. This small amplitude spin fluctuation can induce 
the intra-band transitions of the excited electrons to lower and lower
energy state.  
It is noted that the total energy decreases due to an energy dissipation 
originate from the Gilbert damping.  
The energy distribution shows the gentle spatial dependence 
with the reduced average as shown in Fig.~\ref{jh2b}(g) for 
$T=80$. This means that the electronic wavefunctions 
are rather extended though slightly disturbed by the small tilting of the 
spins.  

\noindent
{\bf Stage (III)}: Relaxation process with interband transition.  
The inhomogeneity developed in the previous stage brings about a 
remarkable localization behavior and derives a dynamic relaxation process with 
the interband transition.   
At around $T$$\sim$220, the electronic and local-spin structures show 
a dramatic change characterized by the large amplitude
motion of a local spin marked by an ellipse in Fig.~\ref{jh2b}(d).  
This motion starts at $T$$\sim$200.  At this time, the excited electrons 
reach the $lowest$ states in the upper energy band as shown by the dotted line 
in the upper panel of Fig.~\ref{jh2b}(a), 
and find the localized place to relax furthermore. 
That is, the localization of the electronic state together with the 
large-amplitude local-spin motion with the polar angle of the order of 
$\pi$ occurs concomitantly. This corresponds to
a pair of the in-gap energy levels split off from the edge of the upper and lower 
energy bands at around $T$$\cong$200,  
as seen in the lower panel of Fig.~\ref{jh2b}(a).  
When the separation of in-gap energy levels is smallest, 
the number of excited electrons decreases rapidly.  
After that, the local spin recovers toward its original direction, 
and the in-gap states merges again into the upper and lower energy bands. 
As shown in Fig.~\ref{jh2b}(h) for $T$=224, 
the excitation energy is concentrated around this local spin, 
which in turn drives a motion of that local spin.
In the same way, the relaxation dynamics 
with interband transition occurs again at $T$$\sim$300,  
around another site (see Figs.~\ref{jh2b}(a), (e) and (i)).  
This transition through the in-gap states  
reminds us the Landau-Zener mechanism.  
However, a more thorough study given below reveals 
that it is a $resonant$ transition and 
is completely different from the Landau-Zener process.  

\noindent
{\bf Stage (IV)}: Relaxation process to a meta-stable state.  
After $T$$\sim$400, the active motion of the local spins is finished and 
the alignment becomes nearly perfect AF.  
The excited electrons, however, 
remains more than $\sim$0.7 in the upper energy band.  
This meta-stable state continues for a long time within our simulation
(at least up to $T$$\sim$8000).

\begin{figure}[t]
\includegraphics[width=8.0cm,clip]{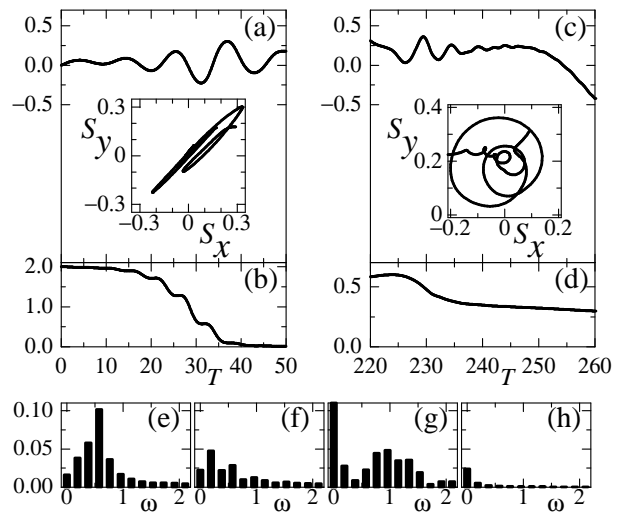}
\vspace{-10pt}
\caption{\label{fft}
Relaxation driven by {\it rapid} oscillation.  
(a) The $y$-component of the local spin which shows
the largest deviation from the AF ground state configuration 
in Stage (I) and (II).  
The inset is the trajectory of the local spin on the $S_x-S_y$ plane.    
(b) The electron occupation number 
at the highest energy states.
(c) The same with (a) but in Stage (III).  
(d) The electron occupation number 
at the lowest energy state (in-gap state) 
in the upper energy band.  
(e) The Fourier spectral weight of (a).  
(f) The same with (e) but $T$=132 in Stage (II).
(g) The Fourier spectral weight of (c).  
(h) The same with (f) but $T$=600 in Stage (IV).
}
\end{figure}

Now we consider the quantum dynamics in more depth.  
As discussed below, there are two components of the 
local spins, i.e., the $rapid$ oscillation and the $slow$ 
motion as expressed by  
${\vec S}_i $$=$$ {\vec S}_i^{slow}$$ + $${\vec S}_i^{rapid}$.  
Let us first consider the $rapid$ oscillation. 
Figure \ref{fft}(a) shows the $y$-component of the local spin moving with 
a largest deviation from the AF ground state configuration 
through Stage (I) and the early period of Stage (II), 
and the inset is the trajectory of the local spin on the $S_x$$-$$S_y$ plane.  
As seen in Fig.~\ref{fft}(a), 
the local spin shows an oscillation with a period of $T_p$$\cong$12, 
i.e., the frequency $\omega$(=$2\pi/T_p$)$\cong$0.5.   
We find that this frequency $\omega$ corresponds to 
the difference of the energy  
between the highest $(\varepsilon_1)$ 
and second highest $(\varepsilon_2)$ energy levels 
in the lower panel of Fig.~\ref{jh2b}(a). 
The rapid oscillation ${\vec S}_i^{rapid}$ is driven by 
time dependence of 
${\vec \sigma}$ in the LLG equation:   
When the
wavefunction has the form 
$|\psi(t)\rangle$$ = c_1 (t)| 1\rangle$$ + c_2 (t)|2\rangle$
with $c_a(t) = c_a(0) e^{-i \varepsilon_a t}$, 
$\langle{\vec \sigma}_i(t)\rangle$ has the component proportional to 
$c_1(t)^* c_2(t) $$\langle 1| {\vec \sigma }_i |2\rangle$
$ \propto e^{ i (\varepsilon_1- \varepsilon_2)t}$ 
and its complex conjugate.  Putting this into the LLG equation, we obtain
${\vec S}_i^{rapid}(t) \propto c_1(t)^* c_2(t) $
$\langle1| {\vec \sigma }_i |2\rangle\times {\vec S}_i^{slow} + h.c.$.
This interpretation is consistent with the 
Fourier spectral weight~\cite{fourier} of the spin motion in Fig.~\ref{fft}(e), 
where the peak is observed around the frequency $\omega \cong 0.5$, 
which corresponds to $\varepsilon_1- \varepsilon_2$ in Stage (I).
This $rapid$ oscillation in turn induces the transition between the 
state $|1\rangle$ and $|2\rangle$, analogously to the electron spin resonance (ESR) 
where the 
oscillating transverse magnetic field induces the Rabi oscillation.
Figure \ref{fft}(b) is the time dependence of the electron occupation number 
at the highest energy states at Stage (I)-(II), and it clearly  
shows this Rabi oscillation with the frequency ($\Omega$) determined by 
the oscillation amplitude ($\delta{S}^{rapid}$) of ${\vec S}_i^{rapid}$. 
From Fig.~\ref{fft}(a), we can read the oscillation amplitude 
$\delta{S}^{rapid}$$\cong$0.3, 
so that the frequency $\Omega$ is estimated to be $J_H\delta{S}^{rapid}$ 
$\cong$0.6.  
Therefore, the occupation number will show an oscillation with a period of 
$2\pi/(2\Omega)$$\cong$5.2.  
This oscillation is actually observed as shown in Fig.~\ref{fft}(b).  
With the Gilbert damping, this oscillation is the damped one and 
the occupation number of the lower energy state increases.  

It is important to note that the spatial inhomogeneity essentially occurs 
with the relaxation dynamics discussed above.  
At Stage (I), the local spin alignment is almost perfect AF.    
In this state, the momentum is a good quantum number and 
the eigenvalue and eigenstate are well defined by it.  
With the time evolution, the excited electrons are distributed into 
lower energy levels, so that the spatial inhomogeneity appears   
as observed in the spatial distribution of the local energy density~\cite{foot2}.  
Reflecting the inhomogeneity, the magnitude of the $rapid$ oscillation 
of the local spin depends on the sites in real space strongly.  
When we look at the local spins at other sites, 
this $rapid$ oscillation is almost missing and only the slow 
and small amplitude motion is observed. 
This means that the site-selective lock-in of
the $rapid$ oscillation of the spins occurs self-consistently 
with the electronic levels before and after the 
transitions. This self-organized space-time structure is the most basic 
mechanism of the quantum transitions in the spin-electron coupled
systems.

With further time evolution in Stage (II), 
the energy level structure becomes disordered reflecting the
disordered spin configuration, and the spectral denstiy does not show
characteristic frequency in this case (see Fig.~\ref{fft}(f)) since it is given by many 
components corresponding to various $\varepsilon_n- \varepsilon_m$.  

Figure \ref{fft}(c) shows the motion of the local spin 
marked by an ellipse in Fig.~\ref{jh2b}(d) for Stage (III).    
As seen in the inset of Fig.~\ref{fft}(c), the local spin shows 
a $rapid$ precession.  
Figure 2(d) shows the electron occupation number 
at the lowest energy state (in-gap state) 
in the upper energy band.  
In this case, however, the Rabi-oscillation behavior 
has not been observed.    
This is because the excited electrons in the states forming the bottom of the upper energy band 
show a cascade relaxation process, 
and the several frequencies are involved in those (see Fig.~\ref{fft}(g)).  
It is seen, in fact, that at the early period of Stage (III), 
the excited electrons occupy the lowest energy states of the upper energy band, 
and from the those states an in-gap state appears (see Fig.~\ref{jh2b}).    
Corresponding to the differences between those energy levels, 
the spectral density in Fig.~\ref{fft}(g) has the peak around $\omega \cong 1$. 
In other words, the inter-band transition occurs successively 
through those energy levels.  
Therefore, the dynamics similar to the ESR process is also essential 
for the inter-band transition in this Stage (III). 
In Stage (IV), there occurs no quantum transition any more
because $\langle 1|\vec\sigma_i|2\rangle$ 
$\times {\vec S}_i^{slow}$$=$$\vec0$ for the AF state.

\begin{figure}[t]
\begin{tabular}{cc}
\begin{minipage}{0.65\hsize}
\includegraphics[width=58mm,clip]{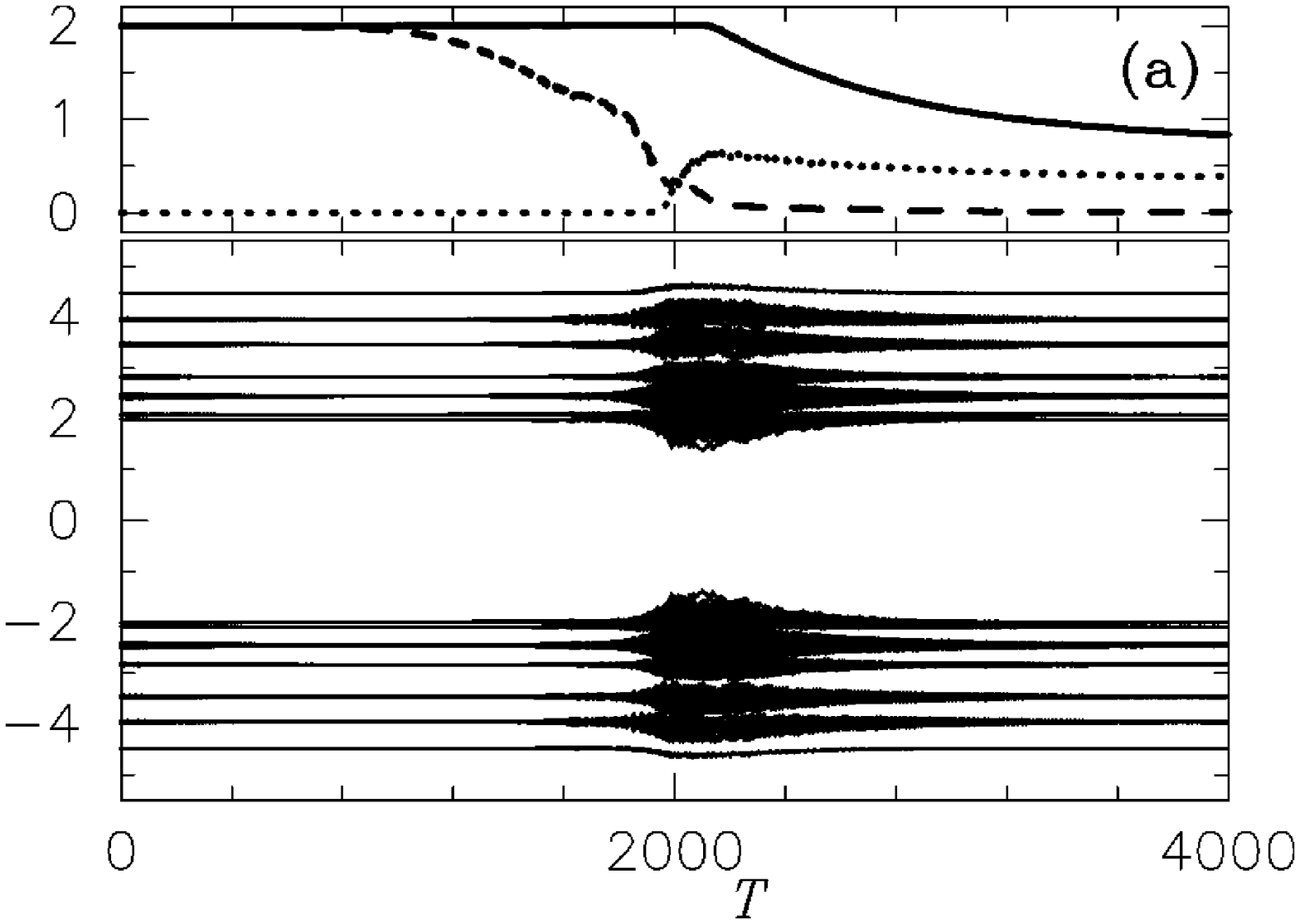}
\includegraphics[width=55mm,clip]{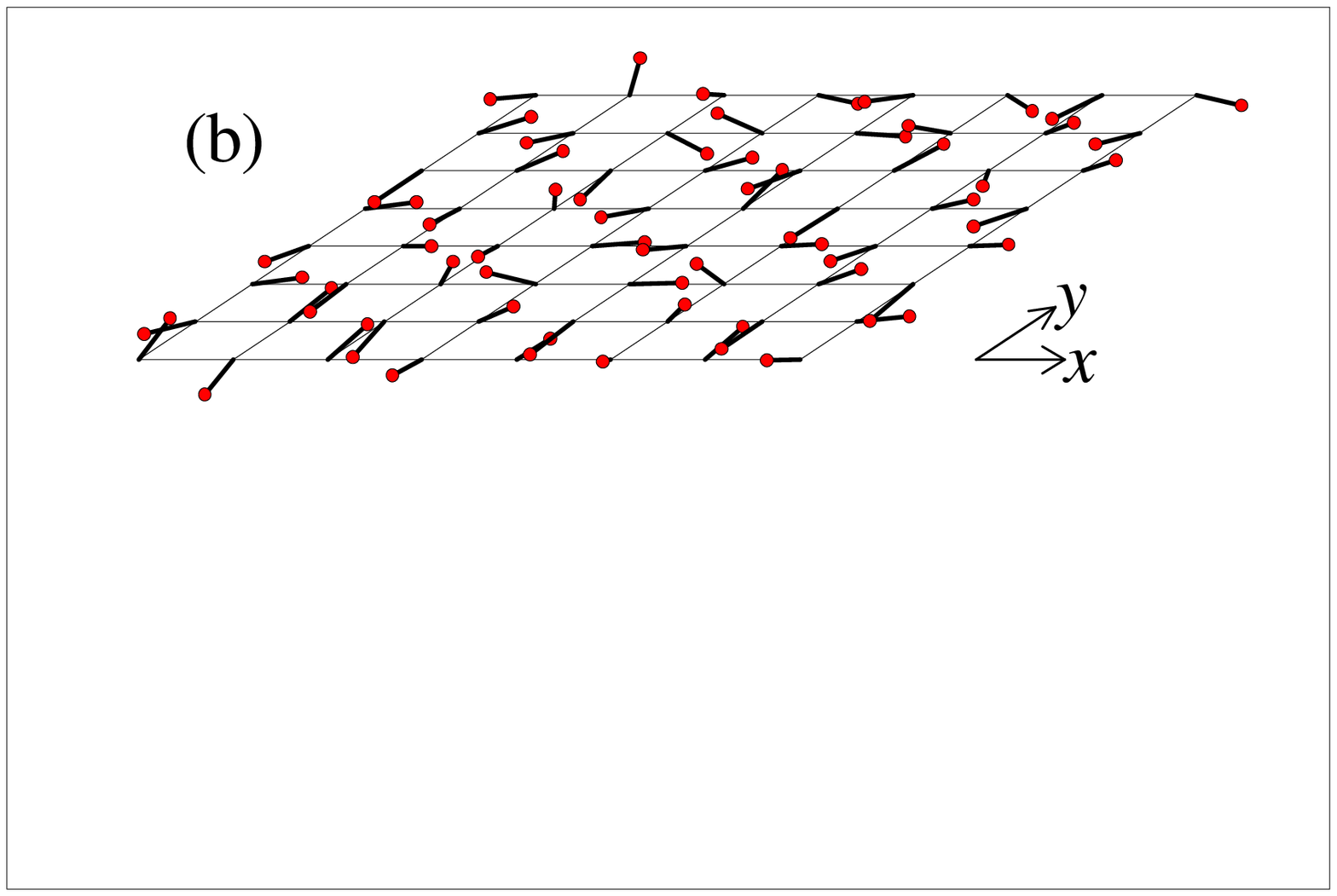}
\end{minipage}
\begin{minipage}{0.3\hsize}
\includegraphics[width=30mm,clip]{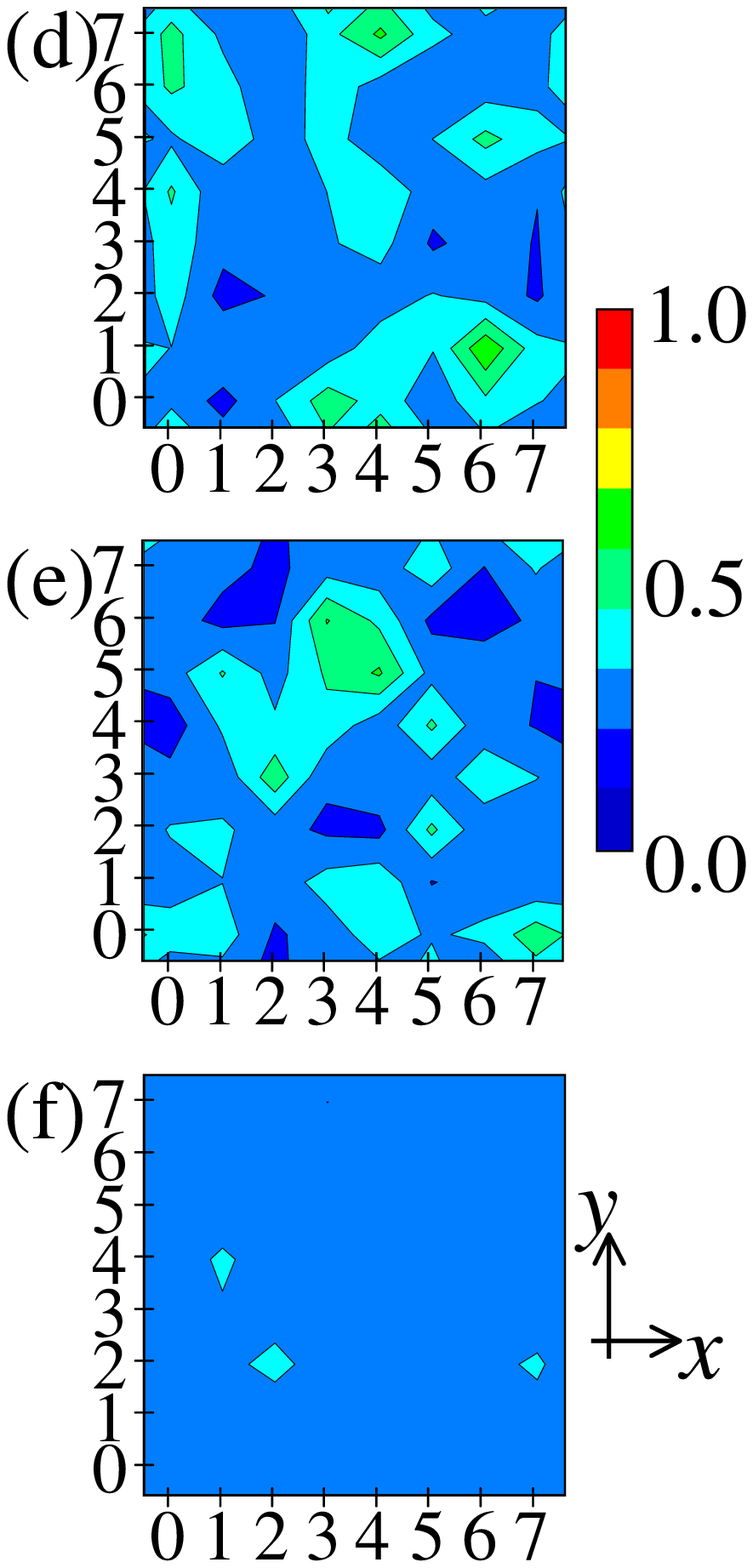}
\end{minipage}
\end{tabular}
\vspace{-10pt}
\caption{\label{a001}
(Color online) Time evolution of the double exchange model.  
The parameter set, 
$t$=1, $SJ_H$=2, $S\alpha$=0.01, $S$=1, $\Delta T$=0.008, is used.
(a) The same with Fig.~\ref{jh2b}(a).  
(b) Structure of local spins at $T$=2000.  
Distribution of energy density at (d) $T$=2000,  
(e) $T$=2100, and  
(f) $T$=2200.  
}
\end{figure}

Let us consider the time scale for the $slow$ motion of the local spins.  
The deviation of the electron spin from the ground state 
of the excited electronic state 
gives a force to the local spins through the LLG equation.  
Therefore, 
$\{(1/N_{\rm eff})J_H S\alpha/[1+(S\alpha)^2]\}^{-1}$ 
determines the time scale for the motion of the local spins,
where $N_{\rm eff}$ is the number of the sites in the real space 
over which the wavefunction of the excitation has the finite 
weight (see the solution of the LLG equation).  
For the plane wave states at Stage (I) and (II),
$N_{\rm eff} = N$ (the total number of the sites) and decreases 
as the wavefunction becomes more localized with the time evolution 
as seen in Figs.~\ref{jh2b}(f)-(i). 
In the present case, $S\alpha =1$, a time scale of the $slow$ dynamics being 
of the order of 100 is derived~\cite{foot}.  
We have also studied the more realistic case of $S\alpha$=0.01 (Fig.~\ref{a001}).  
The mechanism of the relaxation dynamics for $S\alpha$=0.01 is essentially 
the same with the previous case.  
The observed time scale of the $slow$ dynamics in Fig.~\ref{a001}(a) 
is about 10 times as large as that of Fig.~\ref{jh2b}(a), whereas 
the above simple consideration on the time scale gives about 50 times longer one 
compared with the case that $S\alpha=1$.  
In the case for small $S\alpha$, because the $friction$ is small, 
a large number of local spins show very active dynamics (see Fig.~\ref{a001}(b)).  
This results that the considerable spatial area (but $not$ the whole area of the system) 
and many in-gap states 
are responsible for the relaxation dynamics in Stage (III) 
(see Figs.~\ref{a001}(a), (d) and (e)), whereas 
the strongly localized in the real space and a few in-gap states 
are available for the relaxation dynamics in the case that $S\alpha=1$ 
(see Figs.~\ref{jh2b}(a), (h) and (i)).   
As a result, the relaxation process is accelerated, 
and the 10 times difference in the $slow$ time scale appears 
although a hundred times difference in the magnitude of $S\alpha$ is in this case.   

In summary, we have studied the relaxation dynamics of the excited state
of the electron-spin coupled systems.   
The self-organized space-time structure of 
the spins and electronic state triggers the quantum transitions,
and the adiabatic approximation does not work here.

The authors are grateful to Y. Tokura, M. Kawasaki, H. Matsueda,  T. Tohyama, 
S. Ishihara, and K. Tsutsui for useful discussions.
This work is supported by Grant-in-Aid for Scientific Research 
(Grant No. 19048015, 19048008, 17105002, 21244053, 18560043, 19014017, and 21360043)
and a High-Tech Research Center project for private universities 
from the Ministry of Education, Culture, Sports, Science and Technology of
Japan, Next Generation Supercomputing
Project of Nanoscience Program, JST-CREST and NEDO.


\begin{thebibliography}{}
\bibitem{Imada}
M.Imada, A. Fujimori, and Y. Tokura,, Rev. Mod. Phys.{\bf  70}, 1039 (1998).

\bibitem{spincross}
Y. Ogawa, S. Koshihara, K. Koshino, T. Ogawa, C. Urano, and H. Takagi,
Phys. Rev. Lett. \textbf{84}, 3181 (2000).

\bibitem{munekata}
Y. Mitsumori, A. Oiwa, T. S{\l}upinski, H. Maruki, Y. Kashimura, 
F. Minami, and H. Munekata, Phys. Rev. B \textbf{69}, 033203 (2004).

\bibitem{koshihara}
S. Koshihara, A. Oiwa, M. Hirasawa, S. Katsumoto, Y. Iye, C. Urano, 
and H. Takagi, and
H. Munekata, 
Phys. Rev. Lett. \textbf{78}, 4617 (1997).

\bibitem{miyano} K. Miyano, T. Tanaka, Y. Tomioka, 
and Y. Tokura, Phys. Rev. Lett. \textbf{78}, 4257 (1997).

\bibitem{fiebig} M. Fiebig, K. Miyano, Y. Tomioka, and Y. Tokura,
Science \textbf{280}, 1925 (1998).

\bibitem{matsueda1} H. Matsueda and S. Ishihara, 
J. Phys. Soc. Jpn. \textbf{76}, 083703 (2007).

\bibitem{matsueda2}
H. Matsueda, A. Ando, T. Tohyama, and S. Maekawa, 
Phys. Rev. B {\bf 77}, 193112 (2008). 

\bibitem{yonemitsu} 
K. Yonemitsu and N. Maeshima, 
Phys. Rev. B {\bf 79}, 125118 (2009).


\bibitem{deGennes} 
P.-G. de Gennes, Phys. Rev. {\bf 118}, 141 (1960). 

 
\bibitem{IH}
M. Imada and Y. Hatsugai, J. Phys. Soc. Jpn. {\bf 58}, 3752 (1989). 

\bibitem{fourier}
For the Fourier analysis, $2^{12}$ points 
with a small time increment $\Delta T$=0.008 in units of $1/t$ are used.  


\bibitem{foot}
We have examined how long it takes to begin the inter-band transition in Stage (III) 
using 6x6, 8x8, 10x10 and 12x12 systems.  In the initial state, two electrons are 
excited from the lowest to highest energy levels.  We have confirmed that 
the period leading up to the inter-band transition increases linearly as a 
function of the system size.    


\bibitem{foot2}
The inhomogeneity has also been observed in the spatial distribution of 
the magnitude of electron spins $\langle|\vec\sigma_i|\rangle$.  

\end{thebibliography}
\end{document}